\documentclass[10pt]{article}
\usepackage{harvard}

\begin{document}

\begin{center}
\textbf{\Large{Probability as a Physical
Motive}}
\end{center}
\vspace{12pt}

\begin{center}
\textit{\textbf{\large{Peter Martin}}}
\end{center}
\vspace{12pt}

\begin{center} Spatial Information Research Centre\\
University of Otago, Dunedin, New Zealand\\ Phone: +64 3 4798316\\
Email: pmartin@infoscience.otago.ac.nz
\end{center}
\vspace{12pt}

\begin{abstract}
Recent theoretical progress in nonequilibrium thermodynamics,
linking the physical principle of Maximum Entropy Production
(``MEP'') to the information-theoretical ``MaxEnt'' principle of
scientific inference, together with conjectures from theoretical
physics that there may be no fundamental causal laws but only
probabilities for physical processes, and from evolutionary theory
that biological systems expand ``the adjacent possible'' as
rapidly as possible, all lend credence to the proposition that
probability should be recognized as a fundamental physical motive.
It is further proposed that spatial order and temporal order are
two aspects of the same thing, and that this is the essence of the
second law of thermodynamics.
\end{abstract}
\vspace{12pt}

 \vspace{12pt}


\section{Introduction}

``Because that's where the money is.'' According to legend from
America's 1930s era of economic depression and gangsters, that was
the pithy response of the notorious outlaw and jail-breaker John
Dillinger to the question, ``Why do you rob banks?'' It's a catchy
wise crack (albeit referring to reckless and foolhardy behavior),
as it boldly suggests an amusingly simple alternative view of
action, or of motivation---a view perhaps not anticipated by the
question as posed. It is a view which implicitly presumes
possibility rather than necessity, in a figure-ground perceptual
switch, from recognition of that which is prescribed by familiar
rules, to recognition of that which has not been ruled out. It is
a switch from explanation as prior cause (``efficient cause'') to
explanation as attraction (``final cause'') \cite{Salthe:2004}.

The significance of what I have called a figure-ground perceptual
switch must be appreciated, since the point of this letter is to
suggest a sort of inverted view of ``motive'', that is, what we
accept as an explanation for the course of events.

It seems to be a natural habit of human cognition, when presented
with a perceptual field of any sort, to distinguish some figure
from some complementary ground. For example, we have all seen
certain black and white images with which we may experience a
perceptual switch as we swap what we recognize as figure with what
we recognize as ground. Similarly, certain composers have created
such effects musically, so that the listener may experience a
perceptual switch as to which voices carry the figure melody and
which stand as the background harmony.

Such considerations may seem to be of little relevance to
objective science, but I suggest otherwise. For example, in
mathematics, proofs by {\it reductio ad absurdum} are commonly
employed; in this case the problem is more tractable when
approached with a sort of inverted view: one supposes the logical
complement of what one wishes to prove, and derives a
contradiction, which implies that the complement of the
(complementary) supposition must be the case. In electronics, it
is customary to think of circuits in terms of applied voltages as
the  ``background'' and resulting currents as the figures of
interest. Perhaps this is so because voltage sources, such as
batteries, are more commonly encountered than current sources. But
both analysis and design can sometimes be facilitated by a
perceptual switch, thinking of currents as ``ground'' and
resulting voltages as figures of interest. Then, for example, one
speaks of a resistor ``generating'' a voltage, by virtue of the
current flowing through it. In theoretical physics, figure-ground
inversions of point of view have been fruitfully assumed with
respect to the priority of particle or field, where initially a
particle (by its charge or mass) was thought to produce a field,
but later the idea arose that ``particles'' were actually effects
of fields.

In general, science is like puzzle-building, proceeding as far as
possible from one direction, then continuing from another
direction. Impasses provide the signal, not that nature is
essentially paradoxical, but that the explanatory power of the
current direction of work, of the model used to explain nature,
has been exhausted. I suggest that we have reached such a point
with the very model of causal motives as explanations for physical
processes, and that the second law of thermodynamics was the first
signal, the statistical foundations of quantum mechanics the
second signal. Heretofore, causality has been the figure of
interest seen to be operating against a background of uncertainty;
now perhaps we should take the inverted view and accept
uncertainty, or extent of possibilities, as a physical motive, in
the sense that it may account for physical processes with logical
economy and without reference to a particular observer.

So the folk story of John Dillinger is taken as a jumping-off
point for an argument in favor of an alternative model of physical
motive in science, the logical antecedent of which is the second
law of thermodynamics and Carnot's seminal observation that
``wherever there exists a difference of temperature, motive force
can be produced'' \cite{Carnot:1824}. Carnot's observation is
echoed by George Spencer-Brown's statement, on the foundations of
logic: ``There can be no distinction without motive, and there can
be no motive unless contents are seen to differ in value''
\cite{Spencer-Brown:1969}. The role of distinction, or difference,
in both physical and mental processes, is so implicit as to be
ignored, yet it should be recognized as the most fundamental
concept in our models of dynamics. Carnot could as well have said
that where a difference of height exists ({\it e.g.} hydraulic
head), motive force can be produced, or that where a difference of
wind speed exists, motive force can be produced. A generalization
of Carnot's statement of the second law of thermodynamics is that
wherever a difference exists, a motive force also exists, in the
form of the probability that the difference will evolve toward
equilibrium.

I equate differences, in the foregoing broad sense, with
improbability, because there are fewer ``easy'' (low-energy) ways
that states of difference can be realized than states of
indifference can be realized. Hence ``probability'', as used in
speaking of ``probability as a physical motive'', is supposed to
be a measure of opportunities for change of state. In this sense
probability refers more to phase-space volume than to either
ratios of outcomes derived from sampling or to any observer's
degree of certainty as to present or future state.

The proposal is not that this probability should somehow be
sensible to a system, so as to operate as a familiar causal motive
(as one might understand ``pulling'' in terms of the front face of
a hook {\em pushing} on the back face of a hitch); rather the
proposal is that this probability might be recognized as a motive
{\it per se}, insofar as it describes the correlations between
observed states. Given the habit of insistence on proximate,
causal explanation, to assume this view may require what I have
called a figure-ground perceptual switch. But such a view is no
more strange than the natural expectation that ripples on the
surface of a pond should spread outward.

Understanding that any difference or gradient inheres the
potential to produce motive force and {\it vice versa} facilitates
the understanding of such phenomena as, for example, the flight of
albatrosses, which utilize the potential available in wind shear
to stay aloft, or the formation of placer deposits, where stream
flow effectively differentiates alluvium according to density.
Thus distinction (difference, or gradient) and motive are
reciprocally dependent, and this motive, arising from the
possibility of change (or ``flow'') toward equilibrium, is indeed
the most fundamental physical motive.

What is proposed here is not a scientific theory or hypothesis in
the Popperian sense, but a geometric conceptual model of notions
of spatial order and temporal order ({\it i.e.} evolution),
inspired by such useful principles as that of Maximum Entropy
Production (see, for example, \cite{KleidonLorenz:2005}) and
``MaxEnt'' \cite{Jaynes:1957,Dewar:2005}, and by the idea that
probability figures more fundamentally than causality in
theoretical physics \cite{Anandan:2003}. The evolution of life and
biological diversification seem particularly persuasive that
``nature loves opportunities'', leading to Kauffman's conjecture
that living systems ``expand the dimensionality of the adjacent
possible as rapidly as possible'' \cite{Kauffman:2000}. By this it
is meant that evolution proceeds in such a way as to maximize the
number of directions in which life might continue to
evolve---which can be interpreted as another sort of
maximum-entropy principle when entropy is understood as a measure
of ``spread''. The aim of this letter is to synthesize disparate
conceptual models of dynamics and to contribute to the conceptual
foundations of scientific inquiry and theory in this area.

\section{Explanation in Science}

One might suspect that the Dillinger character resonated with
popular imagination because he represented a sort of subversive
freedom at a time when most people's experience was dominated by
severe economic constraints and scarcity of opportunities. But the
idea suggests itself, after the ``because that's where the money
is'' remark, that a similar figure-ground perceptual switch might
often be appropriate in science, as the ultimate aim is to explain
what we observe as simply and as generally as possible, not as
properly as possible. It has often been observed that traditions
of scientific viewpoint define the questions that are asked as
well as the form of hypotheses that are entertained. The
suggestion here is that it may sometimes be possible to simplify
the understanding of nature by revising one's viewpoint, perhaps
even to see exceptions as rules. Such a figure-ground perceptual
switch is justifiable if it leads to logical economy---fewer
primitive concepts and relations accounting for more
observations---regardless whether it conforms to the traditional,
familiar view.

It may be just as useful to presume that nature moves toward
opportunities, as it is to presume that nature is strictly
compelled by causal laws. As a naturalist, one is more impressed
by nature's opportunism than by its competitiveness. Over sixty
years ago Schr\"{o}dinger remarked that ``physical laws rest on
atomic statistics and are therefore only approximate''
\cite{Schrodinger:1944}. More recently Anandan has argued that
``there are no fundamental causal laws but only probabilities for
physical processes'' \cite{Anandan:2003}. Given that these
probabilities are at least invariant with respect to different
observers, and that they lead to correct predictions, perhaps it
is appropriate to recognize these probabilities as the physical
motive that explains change ({\it i.e.} evolution of state). With
such a view, one does not expect some sort of retroactive causal
motive, nor any ``sense'' on the part of physical systems about
their possible future; rather one identifies the field of
possibilities itself as the motive for change.

The opening line was intended to introduce the idea that nature
has an ``outlaw'' character, which is not to say that nature is
haphazard, ergodic, or entirely unruly. The outlaw character of
nature is the ground that essentially complements its ``law
abiding'' character, which has been the figure of interest for
centuries of modern science in pursuit of mechanistic causal
explanation. In Solomonoff's and Chaitin's formal developments of
inductive inference based on algorithmic information theory
\cite{Solomonoff:1964,Chaitin:1974}, a law, as a symbolic string,
is essentially a summary of many ``observation'' strings. In this
scheme, a law can be thought of as the shortest computer program
capable of generating all those other strings which encode
observations to which the law pertains (and it would not generate
any strings representing contrary observations). Thus the
correlations between observations are explicitly taken as the
substance of laws. Laws, as symbolic strings, are summaries that
describe correlations rather than exhibiting them. In information
theory, such ``decorrelation'' is one of the steps of information
compression. Explanation then can be considered to be information
compression; always correlative but not necessarily causal, it is
time-neutral. For there is no requirement that only those
correlations with the proper temporal relation be admitted for
summary. Even if the world were such that future events determined
past events, the correlations could be summarized as laws, which
might be considered as the reasons, or motives, for sequences of
events. The second law of thermodynamics is of this nature:
describing what we expect to observe in the evolution of systems,
it does not appear to explain the ``cause'' that would motivate
such behavior, but it summarizes correlations between
observations.

\section{The second law of thermodynamics, macrostate and
microstate}

The laws of science describe patterns and rhythms that are
observed in nature \cite{Feynman:1965}, as relations between
objects. The statistical tendency for differences to equilibrate,
for the improbable to lead to the probable, is one such recurrent,
reproducible pattern, whether the interpretation be
phenomenological or epistemological. I allude to the controversy
regarding the interpretation of entropy objectively (for example,
\cite{DenbighDenbigh:1985}) {\it vs.} subjectively (for example,
\cite{Jaynes:1965}).

The work of Clausius, Maxwell, Boltzmann, and Gibbs in statistical
mechanics was motivated by the desire to provide a mechanistic
basis for understanding thermodynamics, and in particular for
understanding such notions as that of entropy, in terms of the
details of molecular motions. The second law appears a bit
peculiar in comparison to other physical laws, in that it refers
not to individual objects but to systems: collections of objects
together with their relations. Moreover, as Gibbs observed, in
statistical mechanics one takes a broader view of systems, by
concerning oneself not simply with the evolution of given systems,
but with the evolution of given {\em distributions} of systems
\cite{Gibbs:1902}. Hence the idea of a {\em macrostate}, as an
ensemble of consistent {\em microstates}. The microstates may
differ in microscopic detail, but they are indistinguishable
macroscopically in terms of the chosen system parameters (such as
temperature, pressure, etc., which are typically averages).

There appears again something of a controversy regarding the
matter of reconciling time-symmetric dynamics and time-asymmetric
thermodynamics. The above-mentioned approach, distinguishing more
or less arbitrarily between macrostate and microstate, is
sometimes called ``coarse-graining''; contraposed to this is the
derivation of thermodynamic time asymmetry via choice of
time-asymmetric equations of dynamics (the program of Prigogine
and the Brussels and Austin schools), sometimes called ``extended
dynamics'' \cite{CastagninoGunzig:1998}. A theorem from Lasota and
Mackey \cite{LasotaMackey:1985}, as referenced in
\cite{CastagninoGunzig:1998} is supposed to render the controversy
moot, but since I do not fully appreciate the mathematics, I
choose here the more intuitively-tractable macrostate-microstate
view to make my argument.

\subsection{Phase Space}

Every possible state of a system can be thought of as a point in
an abstract space having as many dimensions as there are degrees
of freedom for the system in all its detail. For example, in a
simple mechanical system comprising four objects, the degrees of
freedom might be considered to be the positions and the momenta of
the four objects. Each object's position could be specified by
three spatial coordinates, and likewise each object's momentum
vector could be specified by three spatial components. So there
would be $4 \cdot (3 + 3) = 24$ degrees of freedom for this
mechanical system of three objects, and it could be represented by
a point in an abstract space of 24 dimensions. In general the
state of such a system of $N$ objects can thus be thought of as a
point in $6N$-dimensional space. This is the phase space of the
system.

Regarding the state of a system, or of the universe as a whole, as
a point in phase space, change is thought of as movement of the
point in phase space. The probabilistic questions of what state
the system is likely to be in, or what state it is likely to
change to, become abstract spatial questions about the proximity
and sizes of regions in phase space.

\subsection{Macrostate as a region of phase space}

A macrostate may be regarded as a certain region (volume) in phase
space, the interior points of which are the microstates that are
consistent with the parameter values that define that macrostate.
In these terms, Boltzmann expressed entropy $S_B$ in terms of
number of microstates $W$:
\begin{equation}\label{BoltzmannEntropy}
  S_B = k\ln W
\end{equation}
where $k$ is Boltzmann's constant. Entropy, then, is associated
with a {\em measure} (generalized volume) of phase space
\cite{Campbell:1965}. Increasing entropy, being associated with
increasing probability, might therefore be understood in terms of
the path of a point in phase space, from one macrostate region to
another macrostate region of larger volume.

The {\em distribution} of microstates associated with a given
macrostate is expected, by the second law of thermodynamics, to
evolve in such a way that its ``envelope'' occupies a larger and
larger phase-space volume as time goes on. The initial phase-space
volume of the distribution is not to be thought of as expanding,
but rather as becoming more convoluted, as it were
\cite{Penrose:1989}, like dye stirred into a liquid, so as to
occupy a larger phase-space volume {\em in terms of the initial
macrostate parameters}. An equivalent statement of the second law
of thermodynamics is that observed macrostates of lesser
phase-space volume ({\it i.e.} less probable macrostates) tend to
go over into macrostates of greater phase-space volume ({\it i.e.}
more probable macrostates).

On the other hand, time-symmetry of dynamical laws means that the
formulas describing the evolution of states (microstates) are
invertible; this one-to-one correspondence between successive
states implies that the phase-space volume of a given distribution
does not change through time (in accordance with the Liouville
theorem). This seems to be the crux of controversy regarding the
objective {\it vs.} subjective nature of entropy: if entropy is
phase-space space volume, and this volume doesn't change for a
distribution of states evolving through time, then it would seem
that there can be no objective increase in entropy. But in fact
for any given macrostate parameters, the overwhelming majority of
microstates consistent with those macrostate parameters evolve to
microstates lying outside the originally-defined macrostate
region. Why? ``Because that's where the space is.''

\section{Prediction {\it vs.} retrodiction}

We are accustomed to thinking that the past is known, whereas the
future is unknown. But surprisingly, if the past is not recorded,
the opposite is more nearly true: it is easier to predict from a
given macrostate into the future, than to retrodict from a given
macrostate into the past. For if past macrostates pass into larger
future macrostates (the second law), while past distributions
maintain their phase-space volume into the future (the Liouville
theorem), then the correspondence between macrostates from past to
future must not be one-to-one; there must be convergence of
macrostates forward in time, specifically toward a macrostate of
equilibrium, and divergence of macrostates backward in time.
Therefore macrostate prediction is easier than macrostate
retrodiction: given a present macrostate, we see that any number
of prior macrostates (of lesser phase-space volume) could have
evolved into the present macrostate, whereas the present
macrostate (along with, perhaps, any number of distinct
macrostates) can be expected, by the second law, to evolve into
the same posterior macrostate of larger phase-space volume. Hence
macrostate prediction is easier than macrostate retrodiction
insofar as the second law of thermodynamics guarantees that
macrostates go over into larger-volume macrostates, even as the
associated microstate {\em distributions} maintain their
phase-space volume as they evolve and their envelopes spread out
into larger macrostates. However, more will be said about this in
connection with Maximum Entropy Production and probability slopes.

Given a particular {\em microstate} on the other hand, prediction
and retrodiction are on equal footing as they employ the same
time-symmetric dynamical laws; thus if one had detailed knowledge
of molecular positions and momenta after an irreversible process
had occurred (such as when a glass of water falls from a table,
breaks, and spills its contents), one could ``run the movie
backward'' to reconstruct the history, the information being
present implicitly in the correlations of microscopic detail. That
is to say, the ``randomized'' microstate after the spill is
actually quite unique when regarded in all its detail, even though
it appears to be a member of a much larger class, in terms of
macrostate parameters, than it had been before the spill. A
complete description of the latter microstate would be extremely
lengthy, except that certain correlations would be discovered
which would allow this microstate to be summarized, in effect, as
``that state resulting from the fall of a glass from a table''.

The foregoing illustrates that information entropy and
thermodynamic entropy are not identical and do not change with
time in the same way. Information entropy is a concept which can
be applicable to an individual system, in accordance with
algorithmic information theory \cite{Kolmogorov:1968}. But
thermodynamic entropy implicitly applies to distributions of
systems, as mentioned previously in connection with Gibbs' work in
statistical mechanics \cite{Gibbs:1902}. Equivalently, one might
subdivide a single system into subsystems and observe the second
law of thermodynamics operating as a ``flow of entropy''
\cite{Haggerty:1974}, down a hierarchy of scale, or from velocity
coordinates to spatial coordinates. One can be confident, by the
second law of thermodynamics, that certain macrostate parameters
of the system, being statistical averages, will change with time
in a way which reflects a larger number of possible microstates
consistent with the latter macrostate. But the number of
parameters required to specify the macrostate need not change. So
any increase in thermodynamic entropy is not reflected in any
increase in information entropy of the macrostate parameters.

An equivalent expression of the second law of thermodynamics, due
to Penrose, is that systems that have been separated in the past
are uncorrelated. Conversely, after ``randomizing'' interactions,
systems remain correlated. Prigogine's insight was that
``irreversibility is a flow of correlations``
\cite{Prigogine:1962}. My interpretation of this is that
correlations flow from the macrostate scale to the microstate
scale, but are conserved.

Thinking solely in terms of least-biased inference, given a
present state, specified by some macrostate parameters, one would
make the same estimate of past macrostate as of future macrostate:
one would expect an increase of entropy either
way.\footnote{Interpreted physically, this view also has th oddly
provincial effect of making ``now'' an uniquely low-entropy time.}
The resolution of this apparent paradox, according to Penrose, is
that the past was constrained, namely by a condition of
exceedingly low entropy (or high improbability) early in the
history of our universe.\footnote{{\it viz.} near-zero {\it Weyl}
tensor of space-time curvature, providing maximal distributed
relative ``height'' of mass, from which it could ``fall'', forming
stars, which would source the free energy for wind, rain, life,
etc.}  In Penrose's view, this early state of our universe in fact
explains the second law of thermodynamics, if only by taking in
its place an initial state of high improbability.

\section{Information, entropy, and order}

For the purpose of this paper, I would like to use the term
"spatial order" to refer to the improbability associated with low
entropy. I acknowledge that many authors would disagree with the
identification of entropy with disorder, some for example
preferring to identify entropy with ``freedom''
\cite{Brissaud:2005}, but I choose the word ``order'' over the
word ``negentropy'' because linguistically it better fits a
property, or a disposition, than does the word ``negentropy''.
Order, as I propose to use the word, is a measure of
disequilibrium, or of compactness in phase space, in so far as a
given state can be regarded as a member of a relatively small
macrostate class. In my preferred terminology, the algorithmic
information content of a highly ordered state would be relatively
low.

Analogously, I would like to use the term ``temporal order''" to
refer to the improbability of a temporal sequence. For a sequence
of high temporal order, relatively few sequences would fit the
``macro'' description of the sequence, and its algorithmic
information content would likewise be low. I conjecture that
spatial order and temporal order are two aspects of the same
thing, and that from any observer's point of view, they exhibit a
reciprocal relation, as temporal order induces spatial order
(``The flow of energy through a system tends to organize that
system'',\footnote{Quote attributed to R. Buckminster Fuller.} the
``dissipative structures'' of Prigogine, etc.), and spatial order
induces temporal order (The spontaneous organization of a system
tends to accommodate the flow of energy through that system: MEP
\cite{Paltridge:1979,SchneiderKay:1994}).

Gibbs refined the Boltzmann expression of entropy in equation
\ref{BoltzmannEntropy}, to account for the possibility that not
all microstates be equally probable:

\begin{equation}\label{GibbsEntropy}
  S_G = -k_B\sum p_i\ln p_i
\end{equation}
where $k_B$ is again Boltzmann's constant, and the $p_i$ are
members of a probability distribution of microstates consistent
with the macrostate.\footnote{The negative sign appears because
the microstate probabilities $p_i$ are normalized to $\sum p_i =
1$, so all $\ln p_i$ are negative.}

Shannon, working in the field of communications theory
\cite{Shannon:1948}, quantified the uncertainty $U_S$ of a message
yet to be received as:

\begin{equation}\label{ShannonEntropy}
 U_S = - \sum p_i \log_2 p_i
\end{equation}
where the $p_i$ comprise a probability distribution of possible
messages. Clearly equations \ref{BoltzmannEntropy} and
\ref{GibbsEntropy} are isomorphic, being measures of the spread of
distributions of possibility, so by analogy Shannon ventured to
call his measure of information ``entropy''.

Notwithstanding Shannon's own warnings regarding careless use of
such terms as ``information'' and ``"entropy''
\cite{Shannon:1956}, there is much disagreement in the literature
about their relation, in spite of such insights as that of
Landauer and Bennett relating the two \cite{Bennett:1982}. Not to
confuse knowledge with information, I accept the Shannon use of
the term ``information'' to refer to a measure of the capacity to
inform, hence uncertainty. Noting the mathematical isomorphism
between Shannon's information entropy and Gibbs' thermodynamic
entropy, some authors have attempted to construct second-law
analogies between biological ``information'' and thermodynamic
entropy \cite{BrooksWiley:1988}, while others have rejected
information theory and constructed their own definitions of
information \cite{Wicken:1980}. In fact, Wicken's definition of
``information'', based on improbability, is close to my proposed
definition of ``order''.

I believe that disagreements about ``information'' and ``entropy''
have to do with the question of how they change with time. The
mathematical isomorphism can hardly be considered accidental, yet
a strictly parallel second law need not follow. On the contrary,
it is enlightening to note the opposite arrows of time, with
respect to information entropy {\em decrease} vis-\`{a}-vis
thermodynamic entropy {\em increase} with time: in any
irreversible process of communication, information entropy must
decrease \cite{Martin:2006}. This is so because information is a
difference, a relational quantity, being a measure of the
potential to inform.\footnote{As with other quantities of
potential, like voltage, there is the confusing tendency to assume
an absolute ground, and to treat such quantities as absolute.}

\section{The arrow of time}

The second law of thermodynamics is considered to be at once the
most certain of physical laws and also the most perplexing or
intriguing in its implications, as expressed variously by many
eminent scientists and philosophers of science
\cite{Einstein:1940,Eddington:1928,Prigogine:1980}. Evidently its
certainty is accounted for by the fact that certainty itself is,
in a sense, its subject; its intrigue may be accounted for by the
fact that it stands alone (almost\footnote{Decay of the long-lived
kaon particle appears as an unique time-asymmetric
quantum-mechanical process.}) among the laws of dynamics as the
one that essentially confirms the difference between past and
future and that accounts for any sensible change in the world.

A four-hundred-year tradition of scientific inquiry, expressed in
equations of dynamics that are time symmetric, and in relativity
transformations that mix time with space, has perhaps led away
from an intuitive personal perspective of time and toward a
geometric perspective of space-time. The geometric perspective
seems simpler and more objective, so that it is compelling to
suppose that our intuitive perspective of time, in relation to the
world becoming what it is, may be provincial and somewhat
illusory. Yet the geometric perspective of time as a dimension
forces one to confront the question of why one direction in this
dimension should be different from another, or indeed, why any
particular value of this dimension's coordinate (such as t = 0, or
``now'', in the vernacular) should be accorded any special status.
(In other words, why do we have the unequivocal sense of being
centered at the present, if all of space-time has a ``geometric''
existence?) Ilya Prigogine, who spent a lifetime studying
nonequilibrium thermodynamics and developing equations of extended
dynamics that break time-symmetry, was primarily motivated in his
studies by an interest in these questions of time
\cite{Prigogine:2003}.

The second law of thermodynamics, or increasing entropy, is
apparently not the only arrow of time, as pointed out by Popper
with the example of radiating waves \cite{Popper:1956}, but aside
from the psychological sense of time, it seems to provide the most
pervasive objective evidence of irreversibility. In ``The many
faces of irreversibility'', Denbigh argues that irreversibility
(which I take to be synonymous with time asymmetry) is a broader
concept than that of entropy increase, and that its essence seemsnning out'', or expansion in space, either physical or
abstract \cite{Denbigh:1989}. Why do things expand? ``Because
that's where the space is''?

\section{MEP and MaxEnt}

The ideas of non-equilibrium thermodynamics have found application
in various fields, notably biology and meteorology. Maximal or
minimal principles of energy flow or entropy production in biology
have been proposed by many authors
\cite{Johnstone:1921,Lotka:1922a,Schrodinger:1944,Hamilton:1977,Lovelock:1987,Wicken:1987,BrooksWiley:1988,SchneiderKay:1994,Salthe:2005};
Paltridge, among others, advanced the principle of Maximum Entropy
Production (MEP) in meteorological modeling
\cite{Paltridge:1979,Paltridge:2001}.

MEP is the tendency for open systems far from equilibrium to
respond to imposed differences, or gradients, by evolving to a
steady state in which the production of entropy is maximized. The
principle of MEP appears as a regenerative combination of the
effect of difference-induced flow in organizing a system (``making
a difference''), together with flow-induced organization enhancing
the ability of the system to conduct the flow. If the principle is
known to be valid for a system, then it can be used (like the Le
Chatelier - Braun principle of chemistry, or indeed, like the
second law of thermodynamics in general) to predict averaged
parameters of system behavior, without knowledge of the system's
details.

The aim of the so-called ``MaxEnt'' information-theoretical
formulation of non-equilibrium statistical mechanics due to Jaynes
is to represent most accurately and fully the uncertainty about
the state of a system, given the constraints of all that is known
about the system. Based on the Bayesian concept of probability as
a measure of ignorance, rather than on the Monte Carlo concept of
probability as a sampling ratio, it is used to quantify the spread
of possible microstates, given only knowledge of the macrostate.
It is considered by its advocates to be a very generally
applicable principle of scientific inference, to be used to arrive
at a least-biased estimate of any unknown quantity, given some
limited knowledge of the system under observation.

Recently the MaxEnt method has been extended
\cite{Dewar:2003,Dewar:2004,Dewar:2005} to estimate not just most
likely states, but most-likely state-change paths--{\it i.e.}
system evolution. MEP as a thermodynamic principle describing
system evolution is thereby derived, along with other well-known
physical principles such as the Fluctuation Theorem, which
quantifies the likelihood of so-called ``irreversible'' processes
actually proceeding in reverse.\footnote{To wit, the fluctuation
theorem can be expressed by $p(\sigma_\tau )/ p(-\sigma_\tau ) =
\exp (\tau \sigma_\tau / k_B)$, where $\sigma_\tau$ is the entropy
production over an irreversible path of duration $\tau$,
$p(\sigma_\tau )$ is the probability of a system following the
irreversible phase-space path, $p(-\sigma_\tau )$ the probability
of a system following the reverse path, and $k_B$ is Boltzmann's
constant.} It has been almost a century since Paul and Tatiana
Ehrenfest published {\it The Conceptual Foundations of the
Statistical Approach in Mechanics} \cite{EhrenfestEhrenfest:1912},
which clarified certain points of disagreement; now the time seems
ripe for a reconciliation of divergent views on the nature of
entropy and the second law of thermodynamics. I suggest that this
reconciliation might come about by the recognition of probability,
not as ignorance, and not as dice, but as a physical motive.

\section{Probability as topography}

As described above, in the phase-space view of a system, its state
is represented as a single point. System evolution is then a
sequence of points, and it may be possible to visualize the second
law of thermodynamics as the movement of the point, representing
the state of the system, into ever-larger macrostate volumes. Even
if ``macrostate'' and ``entropy'' have no absolute meaning until
we have specified the set of parameters which define the
thermodynamic state of the system \cite{DenbighDenbigh:1985}, it
is nevertheless true that, regardless of choice of macrostate
parameters, according to the second law of thermodynamics we
should expect this movement, on average ({\it i.e.} except for
``fluctuations''), to proceed from smaller macrostate volumes to
larger macrostate volumes.

If one then imagines macrostate volume inversely mapped to an
``elevation'', one obtains a surface topography on which system
evolution is expected, by the second law, to proceed ``downhill''.
This is admittedly nothing but a conceptual device, lacking
mathematical rigor. But it helps draw attention to the question of
why anything flows downhill, why a puddle spreads. What would the
Dillinger answer be? We might say that water flowing downhill
responds to the physical motive of gravitational force. But we
don't know what gravity is (though to preserve the precept of
local action we may say that it is curvature of space-time); in
fact flow of mass in response to a gravitational gradient is just
the same sort of second-law journey, downhill on the imagined
phase-space topography, as is flow of heat in response to a
temperature gradient, there being more opportunities for matter to
be collected---greater probability for such states to be
realized---than for it to remain dispersed, just as there are more
ways for a bunch of molecules of a given velocity distribution
({\it i.e.} average temperature) to be spatially dispersed than to
be spatially segregated by velocity.

This is admittedly a difficult point because, as reflected in the
Gibbs refinement of the Boltzmann formula for entropy (and
analogously in Shannon's refinement of Hartley's measure of
information,\footnote{The Hartley measure of information, or
uncertainty, is $I_H = \ln A$, where $A$ is the number of possible
outcomes. Like the Boltzmann entropy, which is simply proportional
to the number of microstates, Hartley information does not take
into account differing relative probabilities of particular
outcomes.} the probability associated with a given macrostate
depends not only on the number of microstates consistent with the
macrostate, but also on their individual probabilities. The
individual probabilities in turn are presumably functions of their
energy distributions. There is of course energy in gravitational
potential (collective difference in location of mass). If gravity
is taken to be curvature of space-time, then the assertion that
there are more opportunities for matter to be collected than for
it to remain dispersed is equivalent to the statement that there
are more ways for space-time to be ``wrinkled'' than for it to be
flat.

Continuing with the conceptual device of probability topography
derived from phase space, one can then visualize MEP as ``flow'',
not just downhill, but down the steepest gradient. Of course this
is what one might expect intuitively. I expect that the question
of whether it holds universally amounts to the question of how
probability slope depends on choice of macrostate parameters. The
question of retrodiction may also be recast in terms of finding
the steepest probability gradient back---minimum (negative)
entropy production in the reverse time direction. The work of
Dewar \cite{Dewar:2003,Dewar:2005} puts these intuitive notions on
firmer mathematical ground. More importantly, I think, the work
bridges the gap between phenomenological understanding of MEP and
epistemological understanding of MaxEnt, opening the door to
acceptance of probability as a fundamental physical motive.

It is not the intent of this letter to discuss the relative
reality of mental models, or maps, on the one hand, and the
physical world, on the other. Rather the intent is to suggest the
consideration of ``probability''---extent of possibility in phase
space---as a fundament of the physics of change and spatial
differentiation in our model, rather than as a computational tool
or as ``"metadata''.

\section{Opportunity and necessity}

It has been assumed that probability must be interpreted either
``objectively'' as a measure of ratios of outcomes, or
``subjectively'' as a measure of observers' ignorance
\cite{Jaynes:1957,Dewar:2004}. I claim that there is another
interpretation of probability, outside this dichotomy: probability
may be interpreted as a measure of possibility, or of opportunity
for realization. Some have remarked that Monod's work {\it Le
hasard et la nec\'{e}ssit\'{e}} (Chance and Necessity)
\cite{Monod:1970} was written ``in the gloomiest of existential
traditions'' \cite{Wicken:1987}. I suspect that profound questions
about origins and destinations---reasons for being---were mistaken
for a profound sense of estrangement. Monod's awareness of kinship
with all life, and with nature, is amply apparent; if there was a
sense of estrangement it might have been with the odd circumstance
in which we find ourselves, caught between science's presumption
of causal origins and religion's assumption of purposive destiny.
How does one interpret {\it hasard}, or ``chance''? Is the sense
that of risk? of randomness? More likely it is that of fortune, or
of possibility. This is the sense of probability as a physical
motive.

\section{Reciprocal relations}

The beginning of thermodynamics is generally taken to be Carnot's
observation that a difference of temperature can produce motive
force. The practical realization of this principle is of course
the heat engine (the generic term for such as the steam engine and
internal-combustion engine), where a difference in the form of
chemical potential is first discharged via combustion in order to
generate the temperature difference from which motive force can be
produced. Conversely, motive force can produce a temperature
difference, practically in the form of a heat pump, or
refrigerator.

Over a century after Carnot, in an early foray into nonequilibrium
thermodynamics, Onsager \cite{Onsager:1931} explained the
reciprocal relations that are observed between pairs of difference
(potential) and flow, such as the {\it Peltier} effect (whereby an
electric current develops a temperature difference) {\it
vis-\`{a}-vis} the {\it Seebeck} effect (whereby a heat flow
develops an electrical potential). The work of Prigogine and the
Brussels and Austin schools in non-equilibrium thermodynamics
carried on to explore the many implications of the reciprocal
relations between differences and flows. The generative, as well
as the dissipative, aspect of the second law of thermodynamics is
now more generally recognized (see, for example,
\cite{SchneiderSagan:2005}).

All these discoveries in non-equilibrium thermodynamics lead one
to venture the view that, in the big picture, ``order'' and its
concomitant ``disordering'' comprise a feedback loop, a
self-determining spiral cascade of differences and flows which
change form but which may exhibit ``strange loopiness'', to borrow
Hofstadter's term \cite{Hofstadter:1979}, in that they refer back
to themselves by inclusion. By analogy to Onsager's specific
reciprocal relations, generic ``order'' may behave as a sort of
``auto-reciprocal'' singleton, insofar as any ``dissipative
structure'' resulting from flow also degrades, conversely
generating another flow. Order (as I've presumed to use the word
both spatially and temporally) is necessarily self referential in
space-time, exhibiting regenerative, self-propagating (even if
self-damping, ``senescent'') effects. Order dissipates, but
dissipation orders, and this holds both spatially and temporally.

\section{Spatio-temporal order}

I have suggested that spatial order and temporal order are two
aspects of the same thing. In the geometric model of space-time,
spatial order (improbability of state) and temporal order
(improbability of evolution) determine each other, in the manner
(if the reader will indulge a sloppy analogy) of the electric
field and the magnetic field, or of a Fourier pair such as
position and momentum. This ``static'', geometric conception of
unified spatio-temporal order is an attempt to integrate the
insights of special relativity\footnote{Special relativity
pertains to the apparent equivalence of uniformly-moving and
non-moving frames of reference, so that the laws of physics should
be covariant with respect to a Lorenz transformation, implying the
non-independence of space and time. General relativity pertains to
the apparent equivalence of gravitation and acceleration.} into
the investigation of the second law of thermodynamics and the
consideration of probability as a physical motive. It follows the
tradition of seeking simplification by assuming
invariance---requiring the proposed concept of order not to depend
essentially on point of view. Hence the notion that time and space
must figure similarly with regard to the structure of
spatio-temporal order.

The proposed conceptual model of spatio-temporal order explicitly
extends in both space and time, since the proposed definitions of
spatial order and of temporal order are based on a notion of
improbability of state or of evolutionary sequence, which makes
sense only for some sort of distribution or sequence that is one
of a class of possibilities. A temporal cross-section of the model
would be an illustration of the second law of thermodynamics, but
the real attraction of the model would be in relating the spatial
and temporal aspects of order, and in providing a more complete
perspective on the asymmetry of time.

This spatio-temporal order must exhibit a ``twist'', or
``handedness'', to reflect the fact that (as far as the evolution
of order is concerned) time has an ``arrow''. But if the model is
at all viable, and unless its mathematical description should
involve non-invertible functions, it would imply that knowledge of
local structure of a region (around ``here'' and ``now'', for
example) should allow extrapolation from here-now in all
spatio-temporal directions, as hinted previously in the suggestion
of a {\em minimum} entropy production principle for {\em
retrodiction} of macrostate. One might expect that the current
macrostate was arrived at by the most unlikely evolution, from the
most unlikely prior macrostate, such that both are consistent with
the model as a whole.

\section{Conclusion}

The main point of this communication is to argue that probability
should be considered to be a fundamental physical motive,
sufficient unto itself for explaining change. Some seventy five
years of work in non-equilibrium thermodynamics, revealing the
constructive aspect of the second law of thermodynamics, together
with relatively recent progress substantiating the connections
between information theory (probability) and thermodynamics, lead
to this opportunity to entertain a new conceptual basis for
understanding systems and their evolution, for posing questions
and formulating hypotheses.

I've tried to substantiate the proposed new view of probability by
way of a topological model of phase-space evolution, in which
entropy increases with progress ``downhill'', and MEP is flow in
the steepest direction. I've assumed a definition of ``order'' in
terms of improbability—relatively small macrostate volume—and
proposed an integration of ``spatial order'' and ``temporal
order'' concepts. It is hoped that the second law of
thermodynamics would then be seen as a partial description of the
structure of spatiotemporal order, and that the structure of
spatio-temporal order would afford further insight into time
asymmetry.

Questions that must be addressed include: How, exactly, is spatial
order to be quantified? How is temporal order to be quantified?
How can they be made commensurable? What, then, is the effect of a
Lorentz transformation on the measure of order? If the proposed
probability topography of phase space can be clearly defined, how
will its slope depend on the choice of macrostate parameters?

\section{ACKNOWLEDGEMENTS}

The author is grateful to Education New Zealand and to the
citizens of New Zealand for financial support in the form of the
NZIDRS award, and to anonymous referees for helpful and
challenging questions, suggestions, and remarks.

\end{document}